\documentclass[prb,twocolumn,aps]{revtex4}
\usepackage{amsmath}
\usepackage{amssymb}
\usepackage{graphicx}
\usepackage{color,soul}
\begin{document}
\title{Inductive van der Waals Force between Two Quantum Loops
}
\author{Kicheon Kang}
\email{kckang@jnu.ac.kr}
\affiliation{Department of Physics, Chonnam National University, Gwangju 61186, 
 Republic of Korea}

\begin{abstract}
We study the van der Waals-London force, which is typically associated with fluctuating electric dipoles in atoms, in a mesoscopic circuit consisting of two inductively coupled superconducting loops. We investigate the {\em inductive} van der Waals-London interaction using both semiclassical and quantum electrodynamic (QED) approaches. The semiclassical model predicts a repulsive interaction due to anticorrelated current fluctuations. In contrast, the QED framework, which incorporates virtual photon exchange, reveals a predominantly attractive force. A key contribution comes from a state-independent two-photon exchange, which is absent in the semiclassical description and undetectable by spectroscopy. 
%Our study introduces a new experimental platform for measuring the van der Waals force between individual artificial atoms via controlled mesoscopic circuits.
%{\color{red} 
Our study introduces a theoretical framework for exploring the van der Waals force between individual artificial atoms via controlled mesoscopic circuits.
%}
\end{abstract}

\maketitle

{\em Introduction-}.
The van der Waals-London interaction between two neutral atoms or molecules is ubiquitous in nature and underlies a wide variety of physical phenomena~\cite{parsegian05}. Microscopically, it results from the interaction of fluctuating electric dipoles~\cite{london37}. 
%Recently, van der Waals (vdW) interactions have been exploited to create novel two-dimensional (2D) materials, also known as van der Waals heterostructures~\cite{geim13}. 
The van der Waals (vdW) force is macroscopic in manifestation. However, measuring it between individual atoms remains unachieved. Although indirect spectroscopic measurements have been achieved between two Rydberg atoms~\cite{beguin13}, directly detecting the force is more challenging.

We note that mesoscopic electronic circuits are an ideal platform for studying van der Waals forces between individual ``atoms'', since artificial atoms can be fabricated on mesoscopic or even macroscopic scales. Their larger dipole moments lead to significantly stronger interatomic interactions (see, for example, Ref.~\onlinecite{you11}). Due to remarkable progress in sensitive measurement technology (see, for example, Ref.~\onlinecite{degen17}), we expect the vdW force between individual artificial atoms to be directly observable. Additionally, electric dipole interactions dominate in real atoms, but artificial atoms can be designed to have dominant magnetic (inductive) interactions.

In this Letter, we study the vdW-London interaction in a system of two inductively coupled quantum loops (see Fig.~1), which can be realized with superconducting condensates. 
The mutual capacitive (dipole) interaction can be reduced through engineering, for example, by spatially separating the two capacitors (Fig.~1(a)) or Josephson junctions (Fig.~1(b)).
Classically, the inductive (magnetic) force between the loops is proportional to the currents, and thus, it vanishes in the absence of current flow. 
This system enables the study of the van der Waals-London interaction, which arises from quantum fluctuations in the current of the zero-current ground state.  We find that the behavior of the inductive vdW-London interaction differs from the standard electric dipole-dominated case. Using a quantum electrodynamic(QED)  approach, we demonstrate that inductive vdW-London interactions exhibit more complex properties, primarily due to the two-photon process of the ``atom''-photon interaction Hamiltonian (the $\mathbf{A}^2$ term, where $\mathbf{A}$ is the vector potential of the radiation field). An interesting geometric feature of the Fig. 1 setup is that the distance $R$ between the loops can be much shorter than their size. This condition is unattainable with real atoms in three dimensions.

%\begin{figure}[l]
\begin{figure}
\centering
\includegraphics[width=8.8cm]{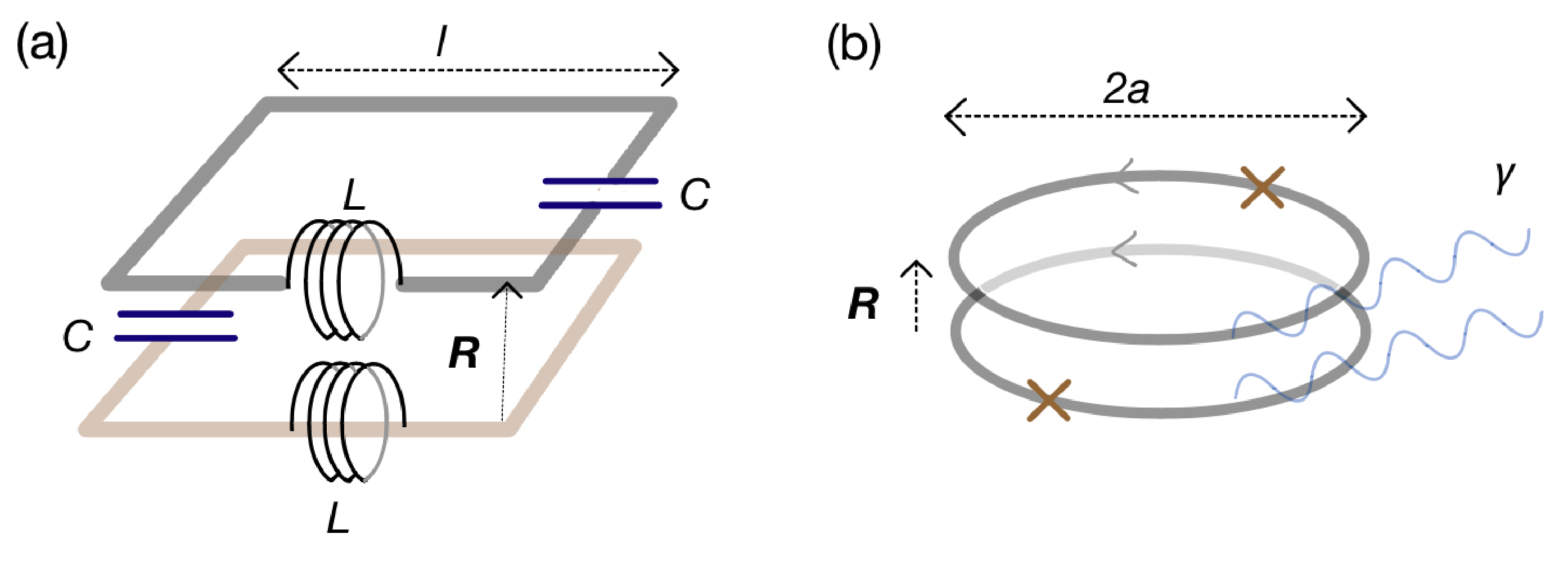} 
\caption{
Schematic diagrams of the two inductively coupled quantum loops. To ensure the inductive interaction dominates, the loops share a close inductive boundary while their capacitive elements are spatially separated: (a) In the semiclassical model, the loops are represented by two inductively coupled quantum LC circuits. (b) In QED, vacuum photons mediate the inductive interaction between the loops, while the Josephson junctions (denoted by X) are positioned far apart.}

\end{figure}

{\em Semiclassical approach with LC circuit model-}.
Two inductively coupled quantum loops (Fig.~1(a)) can be modeled by 
the Lagrangian
\begin{subequations}
\label{eq:L}
\begin{equation}
 {\cal L} = {\cal L}_0 + {\cal L}_i \,,
\end{equation} 
where  
\begin{equation}
 {\cal L}_0 = \frac{L}{2}\dot{Q}_1^2 - \frac{Q_1^2}{2C}
            + \frac{L}{2}\dot{Q}_2^2 - \frac{Q_2^2}{2C}
\end{equation}
and
\begin{equation}
 {\cal L}_i = M\dot{Q}_1\dot{Q}_2  
\end{equation}
\end{subequations}
represent each loop and the mutual interaction, respectively.
Here, $Q_i$ denotes the charge at  circuit $i$ ($i=1,2$), and its time
derivative, $\dot{Q}_i$, is the current.
For simplicity, the two loops are considered identical, each with 
inductance $L$ and capacitance $C$, and the 
angular frequency is $\omega_0 = 1/\sqrt{LC}$.
The mutual inductance,
\begin{equation}
	M(\mathbf{R}) = \frac{\mu_0}{4\pi} \oint_1\oint_2 
     \frac{d\mathbf{r}_1\cdot d\mathbf{r}_2}{|\mathbf{r}_1-\mathbf{r}_2|} ,
\end{equation}
depends on the loop geometry and the displacement vector $\mathbf{R}$.
The Euler-Lagrange equation yields the mutual force:
\begin{equation}
 \mathbf{F}(\mathbf{R}) = \frac{\partial\cal L}{\partial\mathbf{R}}
   = \frac{\partial M}{\partial\mathbf{R}} \dot{Q}_1\dot{Q}_2 \,.
\end{equation}

This classical force between the loops disappears in the absence of current flow. However, in the quantum regime, with macroscopic condensates in each loop, quantum fluctuations of the current play an essential role. 
Using the transformation 
\begin{subequations}
\begin{equation}
 Q_\pm = \frac{1}{\sqrt2} (Q_1 \pm Q_2), 
\label{eq:change_variable}
\end{equation}
the Lagrangian (Eq.~\eqref{eq:L}) becomes
\begin{equation}
 {\cal L} = \frac{L+M}{2}\dot{Q}_+^2 - \frac{Q_+^2}{2C}
          + \frac{L-M}{2}\dot{Q}_-^2 - \frac{Q_-^2}{2C} \,.
\label{eq:L_tr}
\end{equation}
\end{subequations}
%The Hamiltonian can be obtained from the Legendre transformation of 
%\begin{displaymath}
% H = \dot{Q}_+\frac{\partial\cal L}{\partial\dot{Q}_+} + \dot{Q}_-\frac{\partial\cal L}{\partial\dot{Q}_-} - {\cal L} ,
% \end{displaymath}
%which leads to
The corresponding Hamiltonian is
\begin{equation}
 H = \frac{L+M}{2} I_+^2 + \frac{Q_+^2}{2C}
          + \frac{L-M}{2} I_-^2 + \frac{Q_-^2}{2C} \,.
\label{eq:H_tr}
\end{equation}
where $I_\pm = \frac{1}{L\pm M} \frac{\partial\cal L}{\partial\dot{Q}_\pm}$ is the
current variable associated with $Q_\pm$. 
This represents two independent oscillators of the variables $Q_+$ and $Q_-$ 
with the angular frequencies 
\begin{equation}
 \omega_\pm=\omega_0 / \sqrt{1\pm\eta} ,
\label{eq:omega_pm}
\end{equation}
 where $\eta\equiv M/L$.
%Therefore, we can obtain all physical properties from the transformed Hamiltonian (Eq.~\eqref{eq:H_tr}).

The ground state energy from this semiclassical treatment ($E_\mathrm{sc}$) is the sum of the zero-point energies:
\begin{equation}
 E_\mathrm{sc} = \frac{\hbar}{2}\left( \omega_++\omega_- \right) 
     = \frac{\hbar\omega_0}{2} 
       \left[ (1+\eta)^{-1/2} +  (1-\eta)^{-1/2}
       \right] \,.
\label{eq:Esc}
\end{equation}
%Note that $E_\mathrm{sc}$ is greater than the ground state energy of the two noninteracting
%oscillators ($\hbar\omega_0$) for any value of $\eta$. 
%This implies that the interaction is repulsive, 
%and this can be shown explicitly from the mutual force
%\begin{equation}
 %F = \frac{\partial E_g}{\partial R} 
 %  = \hbar\omega_0 \frac{\partial\eta}{\partial R}
 %    \left[ (1+\eta)^{-3/2} - (1-\eta)^{-3/2} \right] \,  
%\end{equation}
%which is always positive.
The energy shift, $\Delta E_\mathrm{sc} = E_\mathrm{sc}-\hbar\omega_0$, is greater than zero for any value of $\eta$, indicating a repulsive interaction.
For $R\gg l$ (where $l$ is the side length of a loop), the value of $\eta$ is much smaller than 1, and $\Delta E_\mathrm{sc}$ approximates to
\begin{equation}
 \Delta  E_\mathrm{sc}  \simeq 3\hbar\omega_0 \eta^2/8.
 \label{eq:DeltaEg}
\end{equation}
This contrasts with the standard vdW-London interaction, which is caused by capacitive (electric dipole) coupling.
In the circuit model, the standard vdW force can be described by a capacitive interaction term~\cite{kleppner94},
\begin{displaymath}
 -\frac{\kappa}{C} Q_1Q_2 \,,
\end{displaymath}
instead of the inductive interaction. The dimensionless coupling constant, $\kappa$, is related to the dipole interaction. For the parallel geometry of the two capacitors, $\kappa=v_c/4\pi R^3$ at large distances, where $v_c$ is the volume of each capacitor. Using the charge variables of Eq.~\eqref{eq:change_variable}, we can find the two eigenfrequencies:
%\begin{equation}
$ \omega_\pm = \omega_0\sqrt{1\pm\kappa}$.
%\end{equation}
Therefore, for a small coupling constant ($\kappa\ll1$), the change in the zero-point energy is given by $-\hbar\omega_0\kappa^2/8$. This gives rise to the well-known attractive vdW-London interaction, which is proportional to $-1/R^6$.

%An interesting question is why the inductive vdW interaction is repulsive, as seen in Eq.~\eqref{eq:DeltaEg}, rather than the typical attractive interaction generated by fluctuating electric dipoles. 
Why does the {\em inductive} van der Waals-London interaction manifest as a repulsion? Phenomenologically, it can be explained by current correlation. From
 $I_1 = (I_++I_-)/\sqrt{2}$ and $I_2 = (I_+-I_-)/\sqrt{2}$, the ground-state current correlation is given by
 \begin{eqnarray}
  \langle I_1I_2 \rangle &=& \frac{1}{2} 
       \left( \langle I_+^2 \rangle - \langle I_-^2 \rangle \right) 
    = \frac{1}{4} \left( \frac{\hbar\omega_+}{L+M} - \frac{\hbar\omega_-}{L-M} \right) 
 \nonumber \\
    &=&  \frac{\hbar\omega_0}{4L} \left[ (1+\eta)^{-3/2} - (1-\eta)^{-3/2} \right] ,  
 \end{eqnarray}
which is negative. In other words, the two currents are anticorrelated. Since wires carrying opposite currents repel each other, the interaction is repulsive.

 {\em Quantum electrodynamic approach-}.
 In the quantum electrodynamic (QED) approach, the interaction between two loops is indirect. It is mediated by virtual photons (see Fig.~1(b)). The Lagrangian is given by
% \begin{subequations}
 \begin{eqnarray}
  {\cal L} &=& {\cal L}_0 + {\cal L}_i , \\
    {\cal L}_0 &=&  \frac{L}{2}\dot{Q}_1^2 - V_1(Q_1) + \frac{L}{2}\dot{Q}_2^2 - V_2(Q_2)  
        + {\cal L}_\mathrm{em} , \nonumber \\
    {\cal L}_i &=&  \dot{Q}_1 \oint d\mathbf{r}_1\cdot\mathbf{A} 
                         + \dot{Q}_2 \oint d\mathbf{r}_2\cdot\mathbf{A} \,. \nonumber
 \end{eqnarray}
 %\end{subequations}
 Here, ${\cal L}_0$ includes the kinetic and the potential energies ($V_{1,2}$) of each loop, as well as the  electromagnetic field of the vacuum (${\cal L}_\mathrm{em}$).  The current in each loop interacts with the radiation gauge field, $\mathbf A$. The Hamiltonian, derived via the Legendre transformation and quantization, can be written as follows:
\begin{eqnarray}
 H &=& \frac{1}{2L} 
   \left( P_1 - \oint_1\mathbf{A}\cdot d\mathbf{x} \right)^2  
     + V_1(Q_1) \nonumber \\
  &+&  \frac{1}{2L} 
   \left( P_2 - \oint_2\mathbf{A}\cdot d\mathbf{x} \right)^2  
     + V_2(Q_2) + H_\mathrm{em} ,
\label{eq:H}
\end{eqnarray}
where
\begin{equation}
  P_n = -i\hbar \frac{\partial}{\partial Q_n}
\end{equation}
is the canonical momentum conjugate to $Q_n$ ($n=1,2$), and $H_\mathrm{em}$ represents the vacuum electromagnetic field.
The vector potential
\begin{equation}
 \mathbf{A}(\mathbf{x},t) = 
  \sum_{\mathbf{k},\lambda} \alpha_\mathbf{k} 
  \left[ 
    u_\mathbf{k}(\mathbf{x}) a_{\mathbf{k}\lambda} e^{-i\omega t}
   +u_\mathbf{k}^*(\mathbf{x}) a_{\mathbf{k}\lambda}^\dagger e^{i\omega t}
  \right] \hat{e}_\lambda \,,
\end{equation}
is expanded in plane wave modes
$u_\mathbf{k}(\mathbf{x}) =  e^{i\mathbf{k}\cdot\mathbf{x}}/\sqrt{V}$
with the normalization coefficient of 
$\alpha_\mathbf{k} = \sqrt{\hbar/2\epsilon_0\omega}$ and the corresponding
angular velocity $\omega = ck$.  In the Coulomb gauge, the polarization vector, $\hat{e}_\lambda$, is constrained to the transverse modes 
by the condition $\mathbf{k}\cdot\hat{e}_\lambda = 0$. 
The Hamiltonian can be split as $H = H_0 + H_\mathrm{in}$, where 
\begin{subequations}
\begin{equation}
 H_0 = \frac{1}{2L} P_1^2 + V_1(Q_1) + \frac{1}{2L} P_2^2 + V_2(Q_2) + H_\mathrm{em} , 
\end{equation}
 is the noninteracting part, and the interaction,
$H_\mathrm{in} = W + X$, consists of:
\begin{eqnarray}
 W &=& -\frac{1}{L} \oint_1\mathbf{A}\cdot\mathbf{P}_1 d\mathbf{x}
              -\frac{1}{L} \oint_2\mathbf{A}\cdot\mathbf{P}_2 d\mathbf{x}, \\
 X &=&  \frac{1}{2L} 
   \left(\oint_1\mathbf{A}\cdot d\mathbf{x} \right)^2  
     + \frac{1}{2L}
   \left(\oint_2\mathbf{A}\cdot d\mathbf{x} \right)^2  .
\end{eqnarray}
\end{subequations}
A key difference from the electric interaction is the presence of the two-photon process, $X \propto\mathbf{A}^2$, alongside the single-photon term, $W \propto \mathbf{p}\cdot\mathbf{A}$. Diagrammatic representations are shown in Fig.~2(a). The single-photon process of $W$ can alter the loop state, whereas $X$ does not.  The leading-order interaction for states with current flow is the single-photon exchange via $W$ (Fig.~2(b)), which is equivalent to the inductive (magnetic) interaction in classical electrodynamics. 

\begin{figure}
\centering
\includegraphics[width=7.5cm]{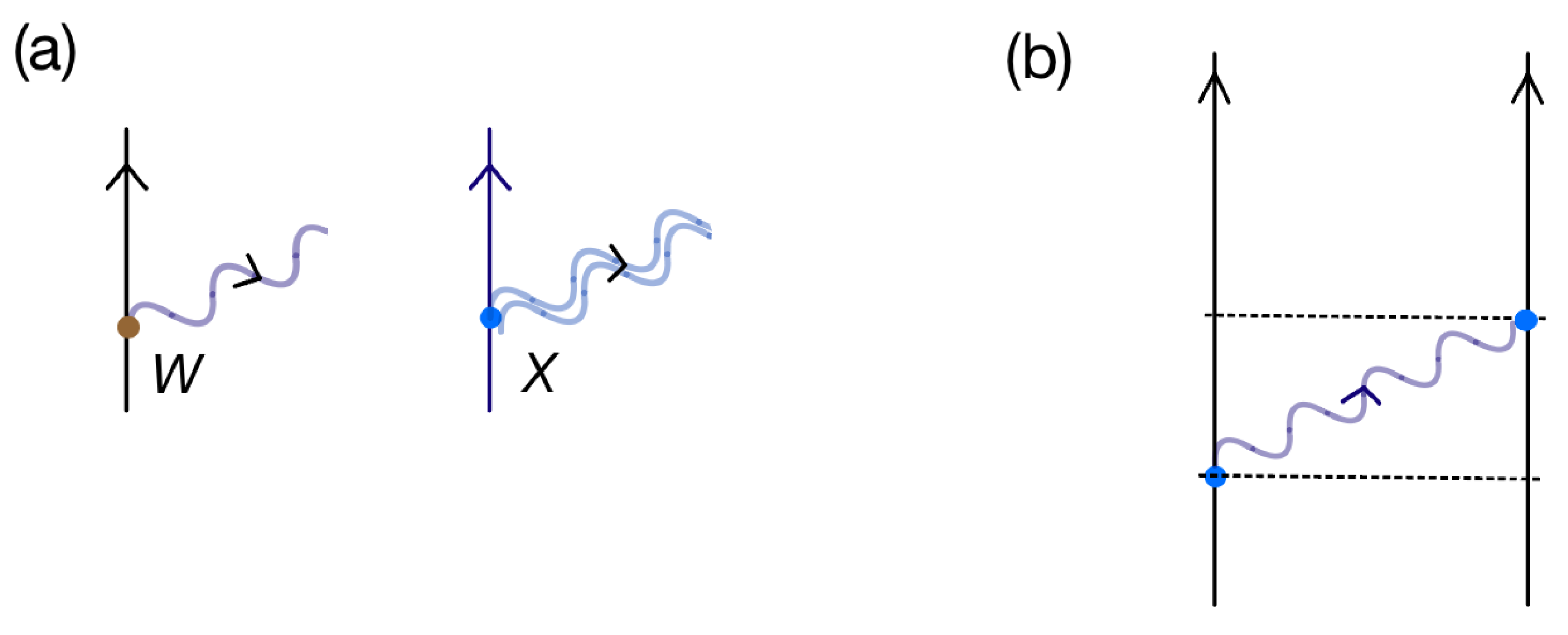} 
\caption{(a) Representation of the interaction between the loop state (solid lines) and the photons (wavy lines). The two types of interaction, $W$ and $X$, involve the single (single wavy line) and double  (double wavy line) photons, respectively.  (b) Leading-order single-photon exchange through $W$ leads to the magnetic interaction between two current-flowing states.
 }
\end{figure}

We calculate the interaction energy in the current-free ground state. The inductive interaction shown in Fig. 2(b) is absent, and the
 leading-order contributions involve two-photon exchange. Due to two types of perturbation, $W$ and $X$,  various terms appear. These terms are depicted schematically in Fig.~3, and the outline of the results from these diagrams is as follows: Here $\Delta E_\alpha$ implies the interaction energy calculated from the process $(\alpha)$. 
 %$\Delta E_\mathrm{i}$ and $\Delta E_\mathrm{i-a}$ are also present in the electric dipole interaction. 
 Process (i) leads to the standard attractive vdW-London interaction, which is proportional to $-1/R^6$ when $R$ is much larger than $2a$, where $2a$ is the diameter of the loop. The term $\Delta E_\mathrm{i-a}$ is related to the retardation effect, also known as the Casimir-Polder interaction~\cite{casimir48}. However, this term can be ignored in our system (see the {\em Discussion} section). 
 %
% It is negligible for $R$ shorter than the crossover length $R_c = hc/\Delta$, where $\Delta$ is the excited state energy of the loop~(see, for example, Sec. II.F.2 of Ref.~\onlinecite{cohen24}).  Physically, this corresponds to the quasi-instantaneous interaction limit, in which the lifetime of the virtual photons is much shorter than the characteristic time ($\hbar /\Delta$) of the loop state. This will be discussed further in the {\em ``Discussion''} section. 
 %
The terms $\Delta E_\mathrm{ii}$ and $\Delta E_\mathrm{iii}$ are unique to the inductive interaction because they include the two-photon interaction of $X$, which is absent in the electric dipole interaction. 
 %The sum of $\Delta E_\mathrm{i}$ and $\Delta E_\mathrm{ii}$ results in a repulsive interaction, equivalent to the semiclassical result for small $\eta$ (Eq.~\eqref{eq:DeltaEg}). The dominant contribution comes from process (iii). This term is independent of the quantum state of the loops and cannot be detected by spectroscopic measurement but governs the mutual force, which gives rise to overall attraction.
 
 %Now, let us calculate each term. 
% {\color{red} 
A detailed derivation of the following interaction energies is provided in the Appendix. Here, we will briefly describe the essential results. %}
 $\Delta E_\mathrm{i}$ consists of a kind of fourth-order interactions in $W$ (see Fig.~3), and is given by
\begin{equation}
  \Delta E_\mathrm{i} = \sum_{a,b,c} \frac{\langle 0|W|c \rangle \langle c|W|b \rangle \langle b|W|a \rangle  
    \langle a|W|0 \rangle}{(E_0-E_c)(E_0-E_b)(E_0-E_a) }.
\label{eq:DEi}
 \end{equation}
 Here $|a\rangle$, $|b\rangle$, and $|c\rangle$ represent the intermediate states in each step of the process. The energy differences are
 $E_0-E_a = -\Delta-\hbar\omega$, $E_0-E_b = -2\Delta$, and $E_0-E_c = -\Delta-\hbar\omega'$, respectively. Eq.~\eqref{eq:DEi} includes self-interaction contributions, which are irrelevant and discarded.  We need to calculate two integrals of the form:
 \begin{equation}
   \frac{1}{V} \oint\oint \sum_\mathbf{k} \frac{ \alpha_\mathbf{k}^2 }{ \hbar\omega+\Delta}  e^{i\mathbf{k}\cdot (\mathbf{r}_1-\mathbf{r}_2)} d\mathbf{r}_1 \cdot d\mathbf{r}_2 .
\label{eq:integral}
 \end{equation}
 This integral is dominated by the contribution with $\omega \lesssim 2\pi c/|\mathbf{r}_1-\mathbf{r}_2|$. At short distances ($R \ll 2\pi\hbar c/\Delta$), the denominator can be approximated as 
 $\hbar\omega + \Delta \simeq \hbar\omega$, which allows us to evaluate the integral analytically.
 This corresponds to the quasi-instantaneous interaction limit (see the Discussion section). Then, we obtain
 \begin{equation}
  \Delta E_\mathrm{i} = -\frac{M^2}{2\Delta} |\bar{I}_1|^2  |\bar{I}_2|^2 ,
 \end{equation}
 where $\bar{I}_n \equiv \langle e| P_n |g\rangle/L$ is the matrix element associated with the current at site $n$ ($n=1,2$).

\begin{figure}
\centering
\includegraphics[width=8cm]{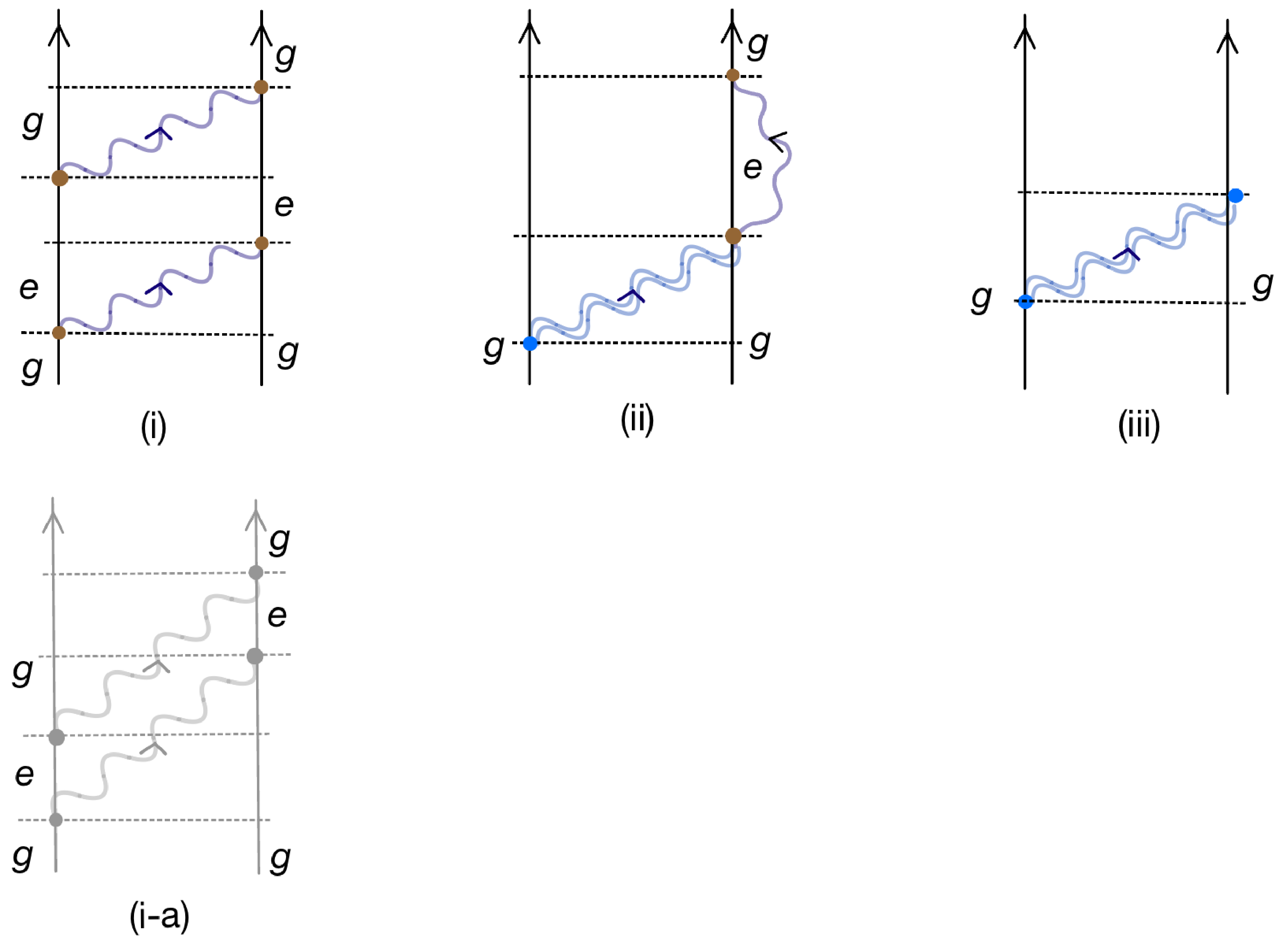} 
\caption{These diagrams illustrate the inductive van der Waals-London interaction between two loops in the current-free ground state via the exchange of two photons. The sum of diagrams (i) and (ii) is equivalent to the semiclassical result. The second-order $X$ process (diagram (iii)) is the predominant contribution. Diagram (i-a), related to the retardation effect, is negligible. Symbols $g$ and $e$ represent the ground and the excited states of a loop, respectively.
 }
\end{figure}

 Process (ii) is peculiar. It combines the single-photon ($W$) and two-photon ($X$) processes. The interaction energy from this process is given by
 \begin{equation}
  \Delta E_\mathrm{ii} = \sum_{a,b} \frac{ \langle 0|X|b \rangle \langle b|W|a \rangle  
    \langle a|W|0 \rangle + \mathrm{h.c.}}{(E_0-E_b)(E_0-E_a) },
\label{eq:DEii}
 \end{equation}
 where $E_0-E_a = -(\Delta+\hbar\omega)$ and $E_0-E_b = -(\hbar\omega+\hbar\omega')$.
 Similarly to $\Delta E_\mathrm{i}$, $\Delta E_\mathrm{ii}$ can be calculated using the quasi-instantaneous interaction approximation, $\hbar\omega+\Delta \simeq \hbar\omega$, in the denominator of Eq.~\eqref{eq:DEii}. Additionally, Eq.~\eqref{eq:DEii} includes an integral of the form 
%{\color{red}
 \begin{eqnarray}
  \langle b |\delta W| 0 \rangle &\equiv& \sum_a  \frac{\langle b|W|a \rangle  
    \langle a|W|0 \rangle}{ E_0-E_a }  \\
  &=& \frac{1}{V} \oint\oint \sum_{\mathbf{k}'} \frac{ \alpha_{\mathbf{k}'}^2 }{ \hbar\omega+\hbar\omega' }  e^{i\mathbf{k}'\cdot (\mathbf{r}_1-\mathbf{r}_2)} d\mathbf{r}_1 \cdot d\mathbf{r}_2 . \nonumber
\label{eq:integral2}
 \end{eqnarray}
% }
 The contributions with $\omega' \lesssim 2\pi c/|\mathbf{r}_1-\mathbf{r}_2|$ dominate this integral, so we can neglect $\hbar\omega$ in the denominator %{\color{red}
  (see Apendix A). %}
 Then, we obtain
 \begin{equation}
  \Delta E_\mathrm{ii} = \frac{L}{2} ( |\bar{I}_1|^2 + |\bar{I}_2|^2 ) \eta^2 .
 \end{equation}
 Of the three major contributions, this is the only positive one. It originates from the third-order perturbation of the ground state. Note that the product of two negative energies, $E_0 - E_a$ and $E_0 - E_b$, appears in the denominator of Eq.~\eqref{eq:DEii}, making the overall sign positive.
 
 Process (iii) is of second order in $X$ and is given by
 \begin{equation}
  \Delta E_\mathrm{iii} = \sum_{\gamma} \frac{\langle0| X |\gamma\rangle \langle\gamma| X |0\rangle + \mathrm{h.c.} }{ E_0-E_\gamma } ,
 \end{equation}
 where $\gamma$ denotes a state with two photons created from the ground state, $|\gamma\rangle = a_{\mathbf{k}\lambda}^\dagger a_{\mathbf{k}'\lambda'}^\dagger |0\rangle$. As in the other cases, the self-energy contribution is discarded. The integral of Eq.~\eqref{eq:integral2} appears in this term. We use the same approximation as in the calculation of $\Delta E_\mathrm{ii}$, $\hbar\omega+\hbar\omega' \rightarrow \hbar\omega'$, and obtain
 \begin{subequations}
 \begin{equation}
  \Delta E_\mathrm{iii} = -\frac{\hbar c}{\pi a} \eta(\mathbf{R}) \lambda(\mathbf{R}) ,
 \end{equation}
 where the two geometry-dependent dimensionless functions are given by $\eta=M/L$ and 
 \begin{equation}
  \lambda(\mathbf{R})  \equiv \frac{\mu_0a}{4\pi L} \oint_1\oint_2 
     \frac{d\mathbf{r}_1\cdot d\mathbf{r}_2}{|\mathbf{r}_1-\mathbf{r}_2|^2} .
  \label{eq:lambda}
\end{equation}
\end{subequations}
%It can be rewritten as
%\begin{equation}
%  \Delta E_\mathrm{iii} = -\frac{\hbar c}{R} \eta(\mathbf{R}) \kappa(\mathbf{R}) ,
 %\end{equation}
 %in terms of the dimensionless functions $\eta(\mathbf{R}) = M(\mathbf{R})/L$, and
% \begin{equation}
 % \kappa(\mathbf{R}) \equiv R N(\mathbf{R})/L .
 %\end{equation}
 
 {\em Interaction energy and force-}.
Now, we will analyze the physical implications of these results. $\Delta E_\mathrm{i} + \Delta E_\mathrm{ii}$ reproduces the semiclassical repulsion.  In the circuit model of the loop state in the QED approach, we find that $\Delta =\hbar\omega_0$ and $|\bar{I}_n|^2 = \hbar\omega_0/2L$. Therefore, we obtain $\Delta E_\mathrm{i} = - \hbar\omega_0\eta^2/8$ and $\Delta E_\mathrm{ii} = \hbar\omega_0\eta^2/2$. This leads to
\begin{equation}
 \Delta E_\mathrm{i} + \Delta E_\mathrm{ii} = \frac{3}{8} \hbar\omega_0\eta^2,
\label{eq:DE1+DE2}
\end{equation}
which coincides with the small-$\eta$ limit of the semiclassical result~\eqref{eq:DeltaEg}. This explains the origin of the repulsive interaction found in the semiclassical approach. The repulsive interaction is primarily due to the greater value of $\Delta E_\mathrm{ii}$ compared to the standard attractive vdW-London interaction energy, $|\Delta E_\mathrm{i}|$.
 
 This is not the end of the story. QED contains an additional term, $\Delta E_\mathrm{iii}$, which the semiclassical treatment misses.  This state-independent term cannot be detected via spectroscopic measurements because the energy shifts are identical for all loop states. They can only be observed via force measurements. In our system, force from this contribution, 
 \begin{equation}
  F_3 \equiv -\partial(\Delta E_\mathrm{iii})/\partial R ,
 \label{eq:F3}
\end{equation}
dominates. In other words,
 \begin{equation}
  |F_3/F_{12}| \sim c/(R\omega_0) \gg 1 ,
 \end{equation}
 in the quasi-instantaneous interaction limit ($R \ll 2\pi\hbar c/\Delta$).
Here, $F_{12} \equiv -\partial (\Delta E_\mathrm{i}+\Delta E_\mathrm{ii})/\partial R$ is the force obtained from the contributions (i) and (ii).

 Fig.~4 shows the plots of (a) the semiclassical ground state energy ($\Delta E_\mathrm{sc}$; Eq.~\eqref{eq:Esc}) and (b) the mutual force ($F_3$; Eq.~\eqref{eq:F3}) calculated from the QED approach, both of which are plotted as a function of the distance between the loops. The semiclassical energy leads to a repulsive interaction, whereas the QED result is predominantly attractive. Both results are meaningful in real experiments. The ground state energy is plotted using the semiclassical result, which is equivalent to the QED result of $\Delta E_\mathrm{i} + \Delta E_\mathrm{ii}$  in the $\eta\ll1$ limit. This result exhibits a repulsive interaction, which is relevant to measurements involving transitions between the different energy levels. In other words, spectroscopic measurements would reveal this repulsive interaction. Conversely, the mutual force is governed by $F_3$, a purely a QED result that shows attractive interactions. This term shifts the energy of all levels equally and cannot therefore be detected by a spectroscopic measurement.

\begin{figure}
\centering
\includegraphics[width=8cm]{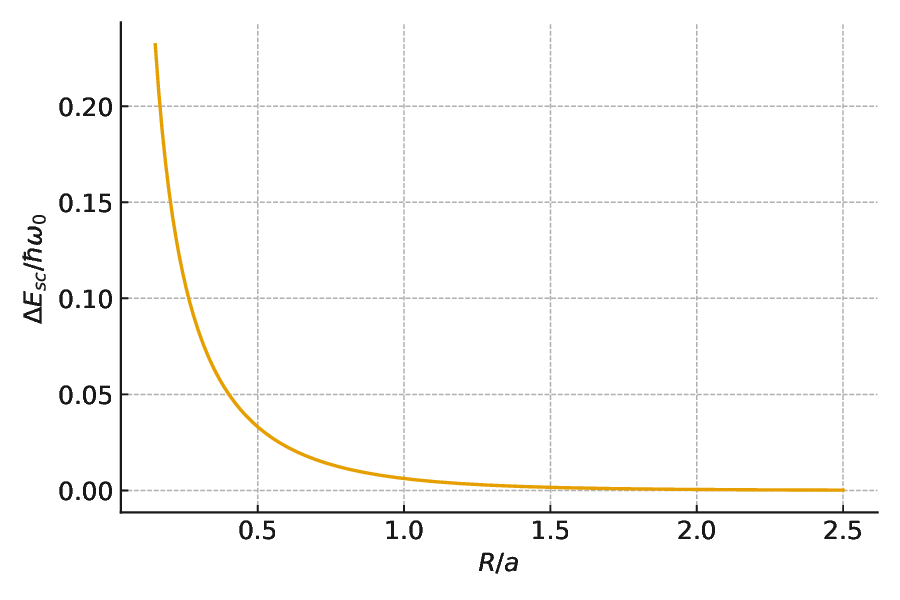} 
\includegraphics[width=8cm]{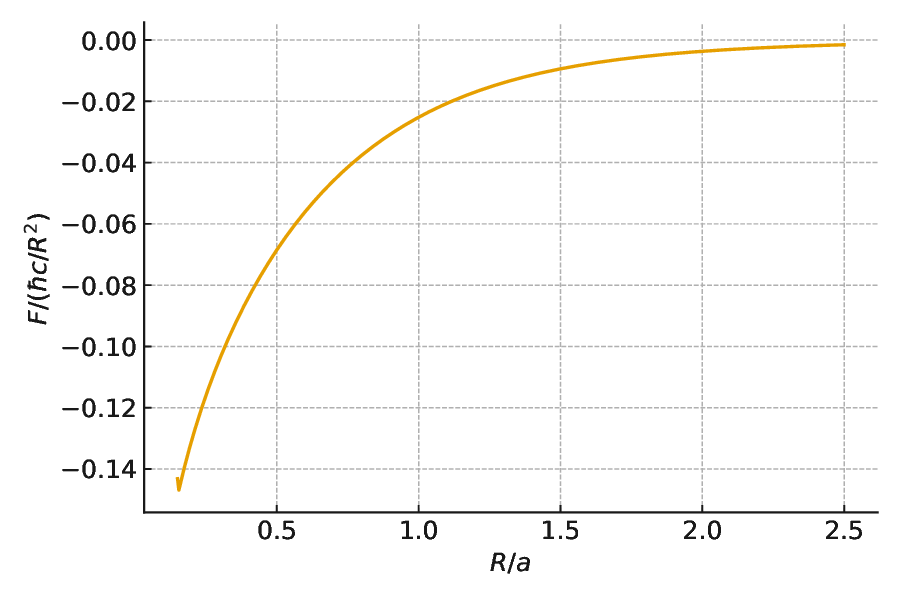} 
\caption{(a) The semiclassical interaction energy, $\Delta E_\mathrm{sc}$ (see Eq.~\eqref{eq:Esc}), and (b) the mutual force, $F_3$ (see Eq.~\eqref{eq:F3}),  calculated from the QED approach, as a function of the distance $R$ between the two circular loops with $b/a=0.05$. Note that the dominant QED term yields an attractive mutual force, $F_3$, which contrasts sharply with the repulsive interaction energy $\Delta E_\mathrm{sc}$, or $\Delta E_i+\Delta E_{ii}$ (see Eq.~\eqref{eq:DE1+DE2}). 
 }
\end{figure}

 {\em Discussions-}.
 %{\sf Long vs. short distance limits} \\
 The interaction energies exhibit different behaviors at the long ($R\gg 2a$) and short ($R\ll 2a$) distances. For $R\gg 2a$, we find that the semiclassical interaction energy is given by $\Delta E_\mathrm{sc} \sim 1/R^6$, which is the typical length dependence of the vdW interaction (except that $\Delta E_\mathrm{sc}$ leads to a repulsive interaction). However, the QED-specific contribution shows a different behavior: $\Delta E_\mathrm{iii} \sim -1/R^7$. The short-distance limit is more interesting because (1) it is more accessible to real experiments, and that (2) it cannot be achieved in real atoms. For $R\ll 2a$, we find that
 \begin{eqnarray*}
  \Delta E_\mathrm{sc} &\sim& \hbar\omega_0 \left[ \ln{(2a/R)} + \ln{4} - 2 \right]^2\\
  \Delta E_\mathrm{iii} &\sim& -\frac{\hbar c}{R}  [\ln{(2a/R)} + \ln{4}-2] .
\end{eqnarray*}
% {\sf UNIVERSAL force at $R\gg 2a$} \\
 The mutual force, dominated by $F_3$, exhibits universal behavior that depends only on $\hbar c/R^2$ multiplied by a geometric factor. For $R\ll 2a$, we find that
 \begin{equation}
  F_3 \simeq -\frac{\hbar c}{R^2} g(R) ,
\label{eq:F3a}
 \end{equation} 
 where $g(R) = [\ln{(2a/R)} + \ln{4} - 1]/ [\ln{(2a/b)} + \ln{4} - 2]^2$ is a geometry-dependent factor of ${\cal O}(1)$. Here, $b$ denotes the cross-sectional radius of the loop wire.
 % Force estimation
 For a typical mesoscopic scale of about $R = 0.1 \mathrm{\mu m}$, the magnitude of the force ($\sim \hbar c/R^2$)  is approximately $3\times 10^{-12}\mathrm{N}$. Current technology can detect this minute force~(see, for example, Ref.~\onlinecite{moser13}).
 
This expression of the force (Eq.~\eqref{eq:F3a}) resembles the Casimir force, except for the geometry-specific logarithmic factor of $g(R)$. In one dimension, in particular, the Casimir force is equal to $-\hbar c/(24R^2)$. 
This suggests a close relationship between the Casimir force and the inductive van der Waals interaction, specifically process (iii) in Fig. 3. It is important to note that the Casimir effect can be derived without reference to zero-point energies, i.e., in terms of quantum forces between charges and currents~\cite{schwinger75,jaffe05}.

% {\sf Why the diagram (i-a) can be neglected...}
We neglected the term $\Delta E_\mathrm{i-a}$ (Fig.~3),
which is associated with the retardation effect, also known as the Casimir–Polder interaction. 
This contribution is comparable to $\Delta E_\mathrm{i}$ and $\Delta E_\mathrm{ii}$ only at large distances, with $R\sim R_c=hc/\Delta$~(see, for example, Sec.~II.F.2 of Ref.~\onlinecite{cohen24}). For a typical mesoscopic loop, 
$\Delta \sim 10^{-5} \mathrm{eV}$, giving $R_c\sim 0.1\mathrm{m}$. At this distance, the typical energy scale is $\Delta E_\mathrm{i-a} \sim \Delta \eta^2 = {\cal O}(10^{-22}) \mathrm{eV}$. This is far below measurable levels. Therefore real experiments could operate in the quasi-instantaneous interaction limit, where this term is negligible.

To fully appreciate the inductive vdW force, it is necessary to contrast our model with natural atoms. In real atoms, the ratio of magnetic to electric dipole moments is strictly constrained by fundamental physical constants, which makes the magnetic vdW interaction inherently much weaker than the standard electric (capacitive) van der Waals force. However, artificial atoms defined by mesoscopic $LC$ circuits are not bound by these fundamental atomic limits. Their effective dipole moments arise from macroscopic charge and current fluctuations within lumped-element components. Therefore, the relative strength of the capacitive coupling ratio ($\kappa$) and the inductive coupling ratio ($\eta$) is a highly engineerable parameter. 

To rigorously describe these systems, our framework naturally generalizes to include the electric dipole interactions arising from mutual capacitance. When both types of interactions are incorporated, the effective interaction Lagrangian for the two coupled loops takes the standard canonical form:
\begin{subequations}
\begin{equation}
 {\cal L}_i = L\eta\, \dot{Q}_1\dot{Q}_2 - \kappa \frac{Q_1 Q_2}{C}  ,
\end{equation}
which modifies the eigenfrequencies of Eq.~\eqref{eq:omega_pm} as
\begin{equation}
 \omega_\pm=\omega_0 \sqrt{\frac{1\pm\kappa}{1\pm\eta}} .
%\label{eq:omega_pm}
\end{equation}
\end{subequations}
Evaluating the total interaction energy under this generalized model reveals that, even in regimes where the couplings are engineered to be comparable ($\kappa \sim \eta$), the inductive effect remains clearly observable. Ultimately, the resulting mutual force is still dominated by the zero-point current fluctuations (the $F_3$ term).

Furthermore, the spatial and energetic decoupling of components in these circuits allows for the active suppression of $\kappa$ independent of $\eta$. This independent tunability is a standard architectural principle in modern circuit quantum electrodynamics. For example, coupled superconducting flux qubits routinely maximize mutual inductance while positioning Josephson junctions on distal segments to explicitly minimize mutual capacitance~\cite{hime06}. Similarly, the ``gmon'' architecture achieves fast, tunable inductive coupling via a shared grounded inductor while employing electrostatic shielding to reduce capacitive crosstalk strictly to zero~\cite{chen14}. Therefore, mesoscopic artificial atoms can be designed to operate in the $\eta \gg \kappa$ regime, cleanly isolating the inductive van der Waals force for experimental observation.
% }

 {\em Conclusion-}.
In summary, we have investigated the van der Waals-London interaction through inductive coupling between two superconducting quantum loops. Our study reveals a fundamental distinction between the semiclassical and quantum electrodynamic treatments of this mesoscopic system. The semiclassical approach, based on the shift in the zero-point energy in coupled LC oscillators, predicts a repulsive interaction due to anticorrelated currents. In contrast, the QED treatment shows the opposite result. The dominant contribution to the mutual force comes from the state-independent two-photon exchange term, $\Delta E_{iii}$, resulting in a predominantly attractive force with a short-distance dependence $F_3 \sim -\hbar c/R^2 g(R)$, reminiscent of the Casimir effect. The key QED term is crucial yet invisible in spectroscopy, governing the mutual force ($|F_3/F_{12}| \gg 1$). The role of vacuum fluctuations in QED has been highlighted through various interesting phenomena such as the Lamb shift, spontaneous emission, and others. Our finding is another interesting manifestation of this and reveals the role of quantum fluctuations in the inductive van der Waals force, specifically the $\mathbf{A}^2$ term in the Hamiltonian. The QED-specific term has a striking effect on the interaction, turning the repulsive force into an attractive one.   
%Our work suggests new experimental avenues for observing the van der Waals force in artificial atoms via controlled mesoscopic circuits. Furthermore, this could be a new kind of experimental test for the validity of QED.
%{\color{red} 
Our work suggests a theoretical framework for observing the van der Waals force in artificial atoms via controlled mesoscopic circuits. Furthermore, this offers a new physical insight into macroscopic manifestations of quantum electrodynamics. %}

\appendix

%{\color{red}
\section{Detailed derivation of the QED interaction energies}

To provide a rigorous derivation of the inductive van der Waals-London interaction energies, we calculate the leading-order QED contributions for the current-free ground state. The interaction between the mesoscopic loops and the vacuum electromagnetic field is governed by the single-photon ($W \propto \mathbf{p}\cdot\mathbf{A}$) and two-photon ($X \propto \mathbf{A}^2$) processes.

A key aspect of this mesoscopic circuit regime is the evaluation of the energy denominators in the perturbation expansion. For standard mesoscopic loop parameters, the relevant vacuum photon energy is on the order of $\hbar\omega \sim \hbar c/R$. Because this photon energy is vastly larger than the characteristic energy level spacing of the loops ($\Delta \ll \hbar\omega$), the lifetime of the virtual photons is extremely short compared to the characteristic dynamics of the circuit. This justifies the quasi-instantaneous interaction approximation, allowing us to approximate the energy denominators as $E_0 - E_a = -\Delta - \hbar\omega \simeq -\hbar\omega$. This is also a key approximation when deriving the standard vdW interaction energy, which corresponds to the quasi-instantaneous interaction limit~\cite{cohen24}.

Below, we provide details on the derivations for the three dominant interaction processes under this limit. Self-interaction contributions are neglected in all of these calculations.

\subsection{Process (i): Fourth-order in $W$}
Process (i) represents the standard fourth-order perturbation involving the exchange of two virtual photons via the $W$ operator. Assuming the initial state $|0\rangle = |g, g'\rangle$, the energy shift involves intermediate states $|a\rangle, |b\rangle, |c\rangle$ representing different photon and loop excitation configurations. 

By defining the matrix element $\bar{I}_n = \frac{1}{L}\langle e|P_n|g\rangle$ ($n=1,2$), the effective interaction can be factored into two parts, $\delta W_\alpha$ and $\delta W_\beta$:
\begin{eqnarray*}
 \delta W_\alpha &=& \sum_{\mathbf{k},\lambda} \frac{\langle b|W|a\rangle\langle a|W|0\rangle}{\hbar\omega} \\
 &=& -\frac{\bar{I}_1 \bar{I}_2}{2\epsilon_0 c^2}  \left[ \oint_1\oint_2 \sum_{\mathbf{k}} \frac{1}{k^2} u_{\mathbf{k}}(\mathbf{x})u_{\mathbf{k}}^*(\mathbf{x}') d\mathbf{x} \cdot d\mathbf{x}' + \mathrm{h.c.} \right] .
\end{eqnarray*}
Using the Fourier transform $\sum_{\mathbf{k}} \frac{1}{k^2} e^{i\mathbf{k}\cdot(\mathbf{x}-\mathbf{x}')} = \frac{1}{4\pi|\mathbf{x}-\mathbf{x}'|}$, the spatial integrals evaluate precisely to the mutual inductance $M(\mathbf{R})$, yielding $\delta W_\alpha = -M(\mathbf{R})\bar{I}_1\bar{I}_2$. 

Combining this with the conjugate term $\delta W_\beta = -M(\mathbf{R})\bar{I}_1^*\bar{I}_2^*$ yields the total energy shift:
\begin{equation}
 \Delta E_{\mathrm{i}} = -\frac{1}{2\Delta} \delta W_\beta \delta W_\alpha = -\frac{M^2}{2\Delta} |\bar{I}_1|^2 |\bar{I}_2|^2 .
\end{equation}
In the identical $LC$ circuit model ($\Delta = \hbar\omega_0$ and $|\bar{I}_n|^2 = \hbar\omega_0 / 2L$), this evaluates to the attractive interaction $\Delta E_{\mathrm{i}} = -\frac{1}{8}\hbar\omega_0 \eta^2$, where $\eta = M/L$.

\subsection{Process (ii): The Combined Process}
Process (ii) is a third-order perturbation involving one execution of the two-photon operator ($X$) and two executions of the single-photon operator ($W$). The interaction energy takes the form:
\begin{eqnarray}
 \Delta E_{\mathrm{ii}} &=& \sum_{a,b} \frac{\langle 0|X|b\rangle\langle b|W|a\rangle\langle a|W|0\rangle + \mathrm{h.c.}}{(E_0-E_b)(E_0-E_a)}  \nonumber \\ 
  &=& \sum_b \frac{\langle 0|X|b\rangle\langle b|\delta W|0\rangle + \mathrm{h.c.}}{E_0-E_b} ,
\end{eqnarray}
where $\langle b|\delta W|0\rangle$ is given by Eq.~\eqref{eq:integral2}.
Evaluating the matrix elements over the vacuum modes under the quasi-instantaneous approximations, $\hbar\omega + \Delta \simeq \hbar\omega$ and $\hbar(\omega+\omega')\simeq \hbar\omega'$, allows us to express the energy shift in terms of the identical geometric double integrals derived above. The second approximation is justified by the fact that the dominant contribution to the integral comes from $\omega,\omega' \lesssim 2\pi c/|\mathbf{x}-\mathbf{x}'|$. This is because contributions at higher frequencies cancel each other out due to random phases~\cite{cohen24}.  This results in:
\begin{equation}
 \Delta E_{\mathrm{ii}} = \frac{L}{2} (|\bar{I}_1|^2 + |\bar{I}_2|^2) \eta^2 .
\end{equation}
For symmetric loops, this yields the positive energy shift $\Delta E_{\mathrm{ii}} = \frac{1}{2}\hbar\omega_0 \eta^2$. The sum of Process (i) and Process (ii) exactly reproduces the repulsive behavior calculated semiclassically via the zero-point energy shift (Eq.~\eqref{eq:DeltaEg}).

\subsection{Process (iii): Second-order in $X$}
The final, purely QED-specific contribution is a second-order perturbation in the $X$ term, which is independent of the quantum state of the loops. The energy shift is:
\begin{equation}
 \Delta E_{\mathrm{iii}} = \sum_\gamma \frac{\langle 0|X_1|\gamma\rangle\langle \gamma|X_2|0\rangle + \mathrm{h.c.}}{E_0-E_\gamma} ,
\end{equation}
where $|\gamma\rangle = |k\lambda, k'\lambda'\rangle$ denotes the two-photon excited state. Applying the quasi-instantaneous approximation to the energy denominator $(-\hbar(\omega+\omega') \simeq -\hbar\omega')$ and utilizing the Fourier relationships yields the double spatial integral:
\begin{equation}
 \Delta E_{\mathrm{iii}} = -\frac{\hbar c}{\pi L^2} \left( \frac{\mu_0}{4\pi} \oint_1\oint_2 \frac{d\mathbf{x}\cdot d\mathbf{x}'}{|\mathbf{x}-\mathbf{x}'|} \right) \left( \frac{\mu_0}{4\pi} \oint_1\oint_2 \frac{d\mathbf{y}\cdot d\mathbf{y}'}{|\mathbf{y}-\mathbf{y}'|^2} \right) .
\end{equation}
The first integral is the standard mutual inductance $M(\mathbf{R})$. The second geometric integral is proportional to the dimensionless function $\lambda(\mathbf{R})$ (Eq.~\eqref{eq:lambda}). It can be evaluated analytically by transforming into Cartesian coordinates: $\mathbf{y} = (a\cos\phi, a\sin\phi, R)$ and $\mathbf{y}' = (a\cos\phi', a\sin\phi', 0)$. By substituting $\theta = \frac{1}{2}(\phi + \phi')$ and $\varphi = \phi - \phi'$, we find
\begin{equation}
 \frac{\mu_0}{4\pi} \oint_1\oint_2 \frac{d\mathbf{y}\cdot d\mathbf{y}'}{|\mathbf{y}-\mathbf{y}'|^2} = \frac{\pi}{2} \mu_0 \frac{1 - \sqrt{1-\beta^2}}{\sqrt{1-\beta^2}} ,
\end{equation}
where $\beta = (1 + R^2/2a^2)^{-1}$. In the asymptotic limit $R \gg a$, this reduces to $\frac{2}{R}M(\mathbf{R})$, ensuring geometric consistency and yielding the final interaction energy.
%}

%\begin{thebibliography}{99}
 \bibliography{references}
%\end{thebibliography}

%\newpage

%
\end{document}